\documentclass[12pt,a4paper]{article}

 \usepackage[dvips]{graphics}
 \usepackage{graphicx}
 \usepackage{afterpage}
 \usepackage{enumerate}
\begin{document}

\title{\bf Fourier analysis of Fe I lines in the spectra of the Sun,
$\alpha$ Centauri A, Procyon, Arcturus, and Canopus}
 \author{\bf V.A. Sheminova and A. S. Gadun}
 \date{}

 \maketitle
 \thanks{}
\begin{center}
{Main Astronomical Observatory, National Academy of Sciences of
Ukraine
\\ Zabolotnoho 27, 03689 Kyiv, Ukraine\\ E-mail: shem@mao.kiev.ua}
\end{center}

 \begin{abstract}
We used spectral observations of Fe I line profiles with a 200
000 resolution to determine micro and macroturbulent velocities
in the atmospheres of the Sun as a star, $\alpha$ Cen A, Procyon
($\alpha$~CMi), Arcturus ($\alpha$ Boo), and Canopus ($\alpha$
Car). Isotropic microturturbulent velocities ($V_{mi}$) and
radial-tangential macroturbulent velocities ($V_{ma}^{RT}$) were
found to be a quite suitable approximation to the velocity field
in the atmospheres of all stars studied except Canopus. The
average velocities $V_{mi}$ and $V_{ma}^{RT}$ are $0.8\pm0.1$
and $2.6\pm0.3$~km/s for the Sun as a star, $0.8\pm0.2$ and
$2.9\pm0.4$~km/s for $\alpha$ Cen A, $0.8\pm0.3$ and
$5.9\pm0.2$~km/s for Procyon, $1.0\pm0.2$ and $4.6\pm0.3$~km/s
for Arcturus. The velocity field in the atmosphere of Canopus
can be described  by an anisotropic radial-tangential
distribution of microturbulence  with $V_{mi}^{RT} = 2.1$~km/s
and anisotropic distribution of macroturbulence with
$V_{ma}^{rad} =17\pm2$~km/s and $V_{ma}^{tan}=1.3\pm1.0$~km/s.
 From Fourier analysis of broadening and shapes
of three spectral lines of Fe I, we have  derived the rotation
velocity $V_e \sin i=3.5 \pm 0.2$~km/s for Canopus.
\end{abstract}

\section{Introduction}
     \label{S-Introduction}
An advantage of the Fourier analysis of stellar spectra is that
it allows the separation of the contributions to the Fourier
transforms from various processes which are responsible for line
broadening. Some examples of application of the Fourier analysis
to the investigations of stellar rotation and astrospectroscopic
turbulence as well as the description of its details and
problems can be found in Gray's book \cite{5}. However, it is not
always possible to realize all the advantages of this technique
in actual practice, as the Fourier transforms of the principal
broadening processes are reliably separated in the high
frequency domain, which can be studied only in observations with
high signal-to-noise ratios and high spectral resolutions.
Nevertheless, the Fourier analysis extended the possibilities
for interpreting the broadening of spectral lines produced by
mechanical motions in stellar atmospheres. While it was found
for solar type stars that line broadening and shifts are caused
by the photospheric overshooting convection and oscillatory
motions  \cite{11,19,40,41}, the question of the nature of motions in
the atmospheres of nonsolar type stars remains open  \cite{14,17,31}.

Our analysis is based on unique spectral observations made by
Dravins  \cite{16}, and its main objective is to determine the
rotation velocity and study the distribution of
micro-macroturbulent velocities in the atmospheres of stars of
various types. With the stars appropriately selected, we are
able to solve the following problems.

1. Follow variations in the velocity field when going from the
Sun to other stars in the H--R diagram. The Sun can be used as a
lest star.

2. Models are available for all stars studied. They are either
theoretical three-dimensional (3D) inhomogeneous models (the Sun
and Procyon  \cite{11,40,41}, $\alpha$ Cen A  \cite{41}) or semiempirical
multicomponent models (Arcturus and Canopus  \cite{17}). Thus the
analysis results can be interpreted more or less reliably and
used to explain the granulation boundary in the H--R diagram
found by Gray and Nagel  \cite{29}. Lines in the spectra of the stars
located to the right of this boundary (cooler stars) have a red
C-shaped asymmetry of the solar type, while lines in hotter
stars display a violet inversion asymmetry. The position of the
boundary approximately corresponds to those stars of luminosity
classes III--V in which convective envelopes adjacent to
atmospheric layers can already form. Thus, a red asymmetry of
spectral lines may suggest that granulation exists on the
stellar surface and the velocity field of the thermal convection
(or photospheric overshooting convection) strongly affects the
line profiles. This conclusion is at variance, however, with the
data for stars of luminosity classes I--II, for which Gray's
granulation boundary and the boundary of convection-induced
envelopes (according to the mixing-length theory  \cite{29}) do not
coincide. At the same time a violet asymmetry of stellar
spectral lines may be interpreted as a prevalence of mechanical
motions other than convection in stellar atmospheres
(oscillations, mass loss, etc.).

3. Four stars in our list (the Sun, $\alpha$ Cen A, Procyon,
Arcturus) have convective envelopes. The photospheric
overshooting convection is of different character in each star,
however, and it is of interest to relate the distinctions to the
astrospectroscopic turbulence parameters. The fifth star, the
supergiant Canopus, lies to the left of the granulation
boundary, it has no convective envelope, and the lines in its
spectrum have a violet inversion asymmetry  \cite{16}. Analysis of its
atmospheric velocity field is of special interest.


\section{Observations }
     \label{ Observation }
We used the spectrograms acquired by Dravins  \cite{16} at the
European Southern Observatory in 1982--1984. A double-pass coude
echelle spectrometer provided a 200 000 resolution at a
dispersion of 1 pm/mm. The instrumental profile is nearly
Gaussian in shape with the FWHM equal to 2.3~km/s. The Nyquist
frequency of observations is on $\sigma _N \approx 0.84$~s/km.
The first zero of the Fourier transform of the instrumental
profile lies at the frequency $\sigma \approx 0.32$~s/km. The
$S/N$ ratio exceeds 300.

%
 \begin{table}[!htb] \centering
 \parbox[b]{14cm}{
 \caption{Basic parameters of the stars studied
 \label{T:1} }
\vspace{0.3cm}} \footnotesize
\begin{tabular}{|llllllll|}
 \hline
   Star  & Sp & $T_{eff}$, K & $\log g$   & $m_*/m_{\odot}$    &$ M_{\rm v}$ &$V_e \sin i$, km/s     & $A_{Fe}^{mod}$     \\
 \hline
 Sun           &G2 V       &5770     \cite{1} &4.44    \cite{1}&1.00&4.38  \cite{1}&1.85  \cite{13}     &      7.55\\
 $\alpha $ Cen A &G2 V &5770  \cite{12} &4.29  \cite{21} &1.08  \cite{15} &4.38  \cite{15} &1.8    \cite{19}    & 7.55\\
 Procyon  &F5 IV--V &6500  \cite{45}&4.04  \cite{45} &1.76  \cite{45}  &2.66  \cite{45} &2.9    \cite{19}    &  7.55\\
 Arcturus  &K1 III       &4260  \cite{38}  &0.90  \cite{38} &0.60  \cite{38} &-0.23  \cite{38}&2.4    \cite{26}     &   6.94  \cite{38}\\
 Canopus  &F0 II            & 7350  \cite{42} &1.80  \cite{42} &8.00         \cite{7} &-4.70      \cite{1}&3.5 This study &7.55\\
    \hline
\end{tabular}
\end{table}
\noindent
The principal characteristics of the stars studied were taken,
for the most part, from paper  \cite{21}. Table 1 gives the effective
temperature $T_{eff}$, gravitational acceleration $\log g$,
relative mass $m_*/m_{\odot}$, absolute luminosity $M_{\rm v}$,
rotation velocity $V_e \sin i$, and iron abundance adopted in
the models $A_{\rm Fe}^{mod}$. The lines selected for the
analysis were unblended or had one unblended wing. The profiles
were not smoothed lest additional distortions appear in the
Fourier transforms. First the observed profiles were
interpolated with a step which varied from line to line within
the range 0.001--0.005 nm (see below). Then the right wing
referred to the local continuum was averaged with the left wing,
and thus the lines were symmetrized. The line center was defined
as the center of gravity of the line core, which comprised 10
percent from the minimum. When one wing had a small blend, it
was corrected according to the shape of the unblended wing. As
soon as the wing reached the local continuum level, the profile
was completed by zeros on both sides. The length of the spectral
region remained constant for all lines in this case (in view of
a specially selected step of wavelength interpolation). As a
result, the Fourier transform resolution was the same for all
lines --- $\bigtriangleup\sigma = 0.009$~s/km.

When dealing with Fourier transforms, one has always to take
proper account of the noises in the observed profiles  \cite{27}. If
$V_e \sin i /V_{ma} > 2$, the information about the parameters
$V_{ma}$ and $V_e \sin i$ resides almost exclusively in the
amplitude of the side lobe, which is located at relatively low
frequencies. These parameters are determined quite reliably.
When $V_e \sin i /V_{ma} < 1$ (this is the case here), all
information about $V_{ma}$ and $V_e \sin i$ resides in the
principal lobe, and this presents additional problems in
separating $V_{ma}$, $V_e \sin i$, and $V_{mi}$. The situation
improves when $S/N > 500$ and a side lobe formed due to
saturation of the absorption line can be seen at higher
frequencies in strong lines. This lobe is very sensitive to the
microturbulence, and the velocity parameters can be separated
more reliably. We measured noise levels for every Fourier
transform of the lines at high frequencies beginning with the
frequency of the first zero of the instrumental profile, where
no signal from the line is observed. The average noise was
$-3.3$~dex. This noise is superimposed on the signal beginning
with the frequency  $\sigma \approx 0.12$~s/km. The first zero
in the transforms for our stars is found, as a rule, between
frequencies of 0.13 and 0.2~s/km. Therefore the analysis of the
principal lobe gives the results for any line, while the side
lobe can be analyzed only for those strong lines in which the
side lobe amplitude exceeds the noise level. To be able to
separate the micro-macrovelocities, we selected only strong Fe~I
lines, with the highest $S/N$ ratio. The equivalent widths of
such lines are 10--20~pm. Table 2 gives these lines together
with their parameters. The oscillator strengths found from the
equivalent widths of solar lines were taken from  \cite{6} or
calculated in the way described in  \cite{6}.
%
 \begin{table}[!htb] \centering
 \parbox[b]{15cm}{
 \caption
{Line list \label{T:2} } \vspace{0.3cm}}
 \footnotesize
\begin{tabular}{|cccc|}
  \hline
 $\lambda $, nm & Multiplet  & $EPL$, eV & $\log gf_W$\\
  \hline
 460.29466      & 39   & 1.48  & -3.14   \\
 536.48801      & 1146 & 4.44  & 0.38    \\
 536.54063      & 786  & 3.57  & -1.41   \\
 536.74755      & 1146 & 4.41  &  0.22   \\
 543.45315      & 15   & 1.01  & -2.12   \\
 633.53378      & 62   & 2.20  & -2.23   \\
 633.68328      & 816   & 3.69  & -0.81   \\
 643.08538      & 62   & 2.18  & -2.46   \\
   \hline
\end{tabular}
\end{table}


\section{Modelling the line profiles}
     \label{Modelling}
The LTE line profiles were calculated with the SPANSAT program
 \cite{4}, and the Fourier transforms of the profiles were calculated
with the FFT program from  \cite{5}. The instrumental profile is taken
into account more exactly by the convolution of the calculated
line profile with the instrumental profile rather than excluding
it from the observed profile. That is why we compare the
observed and calculated profiles and their Fourier transforms
which contain the instrumental profile.

Our calculations are based on the homogeneous stellar model
atmospheres constructed with Kurucz's model grid  \cite{37} with the
parameters $T_{eff}$ and $\log g$ from Table~1.

The turbulent velocity field was simulated within the concept of
micro-macroturbulence. We considered three velocity distribution
variants for the microturbulence. \begin{enumerate}
  \item
Isotropic microturbulence --- a Gaussian distribution with the
most probable value $V_{mi}$. In calculations the microturbulence is
taken into account by the convolution with the line absorption
coefficient.
  \item
Nonisolropic microturbulence --- every element in the radial
flow has a complete set of tangential velocities, and the radial
as well as the tangential velocities have Gaussian
distributions. Their convolution gives a radial-tangential
distribution with the dispersion $V^2_{mi}(\mu) =
(V^{rad}_{mi}\mu)^2 + (V_{mi}^{tan})^2(1 -\mu^2)$, where $\mu =
\cos\theta$, $\theta$ being the heliocentric angle.
  \item
Anisotropic microturbulence --- the radial and the tangential
distributions do not intermingle, and the turbulent flow is
either radial with the area $S_{mi}$ or tangential with the area
$1 - S_{mi}$. The anisotropic microturbulence is included in the
algorithm as a Voigt function in the form of a sum of two
distributions: $H = H(V^{rad}_{mi}\mu)S_{mi} + H(V_{mi}^{tan}(1
-\mu^2)^{1/2})(1 - S_{mi})$.
\end{enumerate}

The macroturbulence was assumed to be in the form of two
distributions.
\begin{enumerate}
  \item
Isotropic macroturbulence --- the star is divided into regions
which contain many macroelements with a Gaussian velocity
distribution and the most probable velocity $V_{ma}$. The
macrovelocity is taken into account in this case through the
convolution of the macroturbulent velocity distribution with the
flux spectrum of a star with no macrolurbulence in its
atmosphere.
  \item
Anisotropic macroturbulence (or the radial-tangential model ---
$V_{ma}^{RT}$) --- the radial and the tangential macroturbulence
given as a sum of two Gaussians: $G = G(V^{rad}_{ma}\mu)S_{ma} +
G(V_{ma}^{tan}(1 -\mu^2)^{1/2})(1 - S_{ma})$, where $S_{ma}$ is
the area fraction occupied by the radial component. In
calculations the macrovelocity is taken into account through the
convolution with the line profile for every position on the
stellar disk.
\end{enumerate}

The rotation velocity $V_e\sin i$ was taken as a constant on the
assumption that the star is spherical and rotates as a solid
body. The rotation was simulated by direct averaging over the
star's disk. The limb darkening was allowed for directly.
Position on the disk was specified by two variables --- the
position angle $p$  and the angle $\theta$ between the center of
the apparent disk and the direction of emerging radiation.
Integration over the disk was performed with the Gauss
quadrature formula with a constant weight function (for $\mu^2$)
and the quadrature formula of the highest trigonometrical
precision (for $p$).

The classical Fourier analysis  \cite{22,23,24,25,26,27,28,43} allows all three
velocities ($V_{mi}$, $V_{ma}$, $V_e\sin i$) to be separated.
The separation scheme, described also in  \cite{5}, was realized for
the analysis of lines in the spectrum of Canopus, whose rotation
velocity was rarely measured before. A value of 15 km/s given in
catalog  \cite{47} was derived in 1953 as a very rough estimate from
half-widths of spectral lines. As to the other stars, their
rotation velocities cannot be determined from the observations
used by us more exactly than it was done earlier (see Table 1).

So, the special calculation scheme realized for four stars
except Canopus was as follows. With known and invariable model
atmosphere, rotation velocity, and damping constant (the
correction factor is unity for all stars), a thermal profile
is simulated with the microturbulence, rotation velocity, and
instrumental profile taken into account. Abundances and
oscillator strengths are initially set as given in Tables 1 and
2. Next the Fourier transform of this simulated profile is
compared to the Fourier transform of the observed profile.
Varying $V_{mi}$, we get the best fit of the simulated first
zero and side lobe to the observed ones, and the principal lobe
amplitude in the simulated Fourier transform must be equal to or
slightly higher than the observed amplitude. If the equivalent
widths of the simulated and the observed lines are not equal, we
make them coincide by varying the parameter $Agf$. In this way,
from the coincidence of the first zero, the side lobe, and the
equivalent width, we determine the microturbulent velocity
$V_{mi}$.

At the second stage we separate the Fourier transform of the
residual profile from the Fourier transform of the observed
profile, the former being the Fourier transform of the
macroturbulence distribution, and then we match it against a
grid of Fourier transforms of macroturbulent velocity functions
calculated for different distributions and dispersions. We
obtain the Fourier transform of the residual profile as the
result of the simple operation:
$m(\sigma)=d(\sigma)/[f(\sigma)i(\sigma)g(\sigma)]$. Here we
introduce the following designations for the Fourier transforms
of profiles: $d(\sigma)$ for the observed profile, $f(\sigma)$
for the thermal profile with microturbulence taken into account,
$i(\sigma)$ for the instrumental profile, $g(\sigma)$ for the
rotation function, and  $m(\sigma)$ for the macroturbulent velocity
distribution. When the macroturbulent velocity is determined,
the calculated and the observed central line intensities must
coincide.

\section{Results}
     \label{Results}

%
 \begin{table}[t] \centering
 \parbox[b]{15cm}{
 \caption
{Fourier-analysis results for five stars \label{T:3} }
\vspace{0.3cm}}
 \footnotesize
\begin{tabular}{|cccccccccc|}
  \hline
 $\lambda $, nm & $d_c$  & $W$, pm & $\log \tau_5$&$V_{mi}$&$V_{ma}^{RT}$&$V_{mi}^{RT}$&$V_{mi}$&$V_{ma}$ \\
  \hline
&&&&&Sun&&&&\\
 536.5 &  0.568 &  7.28  &  -1.7  &  0.9  &   2.3  &   1.5 &    2.3 & 0.6  & 1.6   \\
 536.7 &  0.688 &  14.15 &  -1.7  &  0.8  &   2.2  &   1.5 &    2.3 &    - &  -    \\
 633.5 &  0.598 &  9.90  &  -2.3  &  0.8  &   2.8  &   1.5 &    2.8 & 0.5  & 1.9   \\
 633.6 &  0.582 &  11.34 &  -2.0  &  0.6  &   3.0  &   1.2 &    3.1 & 1.0  & 2.2   \\
    Average &   &        &  -1.9  &  0.8  &   2.6  &   1.4 &    2.6 & 0.7  & 1.9   \\
    Error   &   &        &   0.2  &  0.1  &   0.3  &   0.1 &    0.3 & 0.2  & 0.3   \\
    $V_{rms}$&  &        &        &       &  1.3   &       &    1.3 &      & 1.3   \\
  \hline
&&&&& $\alpha$ Cen A&&&&\\
 536.7  & 0.712  & 16.19&   -1.7 &   0.8  &   2.3  &   1.3& 2.4  &   1.5 &    2.0  \\
 543.4  & 0.811  & 19.59&   -2.8 &   1.1  &   3.5  &   1.5& 3.2  &   --  & --      \\
 633.5  & 0.618  & 10.86&   -2.4 &   0.6  &   3.0  &   1.1& 3.1  &   1.1 & 2.1     \\
    Average   &  &       &    -2.3&    0.8 &    2.9 &    1.3& 2.9 &   1.3 & 2.0     \\
    Error     &  &       &     0.4&    0.2 &    0.4 &    0.1& 0.3 &   0.2 & 0.0     \\
    $V_{rms}$ &  &       &        &        &   1.4  &       & 1.4 &       & 1.4     \\
  \hline
 &&&&&Procyon&&&&                                                                  \\
 460.2 &  0.603 &  10.48&   -2.1 &   1.2 &   6.1  &   1.5 & 6.1 &    1.7& 4.1      \\
 536.4 &  0.488 &  10.09&   -1.7 &   0.6 &   5.9  &   1.5 & 6.1 &    0.8& 4.0      \\
 536.7 &  0.514 &  10.86&   -1.7 &   0.5 &   5.6  &   1.5 & 6.1 &    1.6& 4.1      \\
    Average   & &  &   -1.8&   0.8 &    5.9&      1.5&    6.1  & 1.4  &  4.1         \\
    Error     & &  &    0.2&   0.3 &    0.2&     0.0 &    0.0  &  0.4 &  0.0         \\
    $V_{rms}$ & &  &       &       &    2.9   &      &    3.1  &      &  2.9         \\
  \hline
&&&&& Arcturus&&&&                                                                 \\
 536.4 &  0.707 &  13.17 &  -1.8&    1.4&    4.9 & 2.0 &  4.9 &    1.4&     3.4   \\
 536.5 &  0.671 &  11.30 &  -1.6&    1.0&    4.7 & 1.5 &  4.8 &    1.6&     3.2   \\
 536.7 &  0.727 &  13.60 &  -1.7&    0.7&    4.1 & 1.7 &  4.5 &    1.7&     3.2   \\
    Average   &  &         &-1.7  &  1.0  &   4.6  &   1.7 &    4.7&    1.6&  3.3   \\
    Error     &  &         &0.1   &  0.2  &   0.3  &   0.2 &    0.2&    0.1& 0.1    \\
    $V_{rms}$ &  &         &      &       &   2.3  &       &    2.4&       & 2.3    \\
  \hline
&&&&& Canopus&&&&                                                                \\
 536.4 &  0.384 &  7.45 &   -1.4 &   -- &  -- &  2.1 & 15--0.5& 1.2 & 4.4        \\
 536.7 &  0.426 &  9.00 &   -1.6 &   -- &  -- &  2.1 & 16--0.5& 2.1 & 4.5           \\
 643.0 &  0.210 &  4.95 &   -1.3 &   -- &  -- &  2.1 & 20--3.0&  --  &  --          \\
    Average   &    &  &   -1.4  &      &    &  2.1 &    17--1.3&  1.6& 4.4         \\
    Error     &    &  &    0.1  &      &    &  0.0 &    2--1.0 &  0.4& 0.0         \\
    $V_{rms}$ &    &  &         &      &    &      &    8--0.5 &     & 3.1         \\
   \hline
\end{tabular}
\end{table}


We assumed the radial and tangential components to be equal in
the anisotropic models (for microturbulence as well as for
macroturbulence). The areas occupied by them were also assumed
to be similar. Such assumptions are made usually in the Fourier
analysis of line profiles in stellar spectra because the number
of unknown parameters can be reduced in this case. Equality of
$V^{rad}$ and $V^{tan}$ in nonisotropic microturbulence models
results in an isotropic distribution, and so we shall not use
the nonisotropic model in this study.

Table 3 gives the velocities derived for each line analyzed.
Given also are the observed values of $d_c$ and $W$ (with the
instrumental profile not subtracted) as well as the effective
depths of line formation $\log \tau_5$ calculated with the
depression contribution functions. One can judge from these
parameters how warranted is averaging the velocities over all
lines. No conclusion on velocity stratification in the stellar
atmospheres can be drawn based on the depths $\log \tau_{5,W}$
derived here, since the layer in which a line is formed is much
wider for the star as a whole than for a feature on the stellar
disk. Besides, we do not take into consideration the velocity
gradients in the atmospheres, and the number of lines used in
our analysis is too small. The velocities averaged over all
lines are referred to the depths of formation averaged over all
lines. The values in columns 5 and 6 are the most probable
velocities for the model with the isotropic microturbulence and
the anisotropic macro-turbulence, in columns 7 and 8 for the
anisotropic micro-macroturbulent velocities, and in columns 9
and 10 we give the values obtained in the classical way --- the
isotropic microturbulence was found from equivalent widths and
the isotropic macroturbulence from central intensities.

The table gives also the $rms$ values $V_{rms}$ of
macroturbulent velocity distributions, they can be used for
reducing the most probable values $V_{max}$ for the isotropic
and anisotropic distributions of macroturbulent velocities to a
common system with the aim to compare the results. Recall that
$V_{rms} = V_{max}/\sqrt{2}$ for the Gaussian isotropic velocity
distribution and $V_{rms} \approx V_{max}/2$ for the
radial-tangential distribution  \cite{23}.

  \begin{figure}
\includegraphics [scale=1.2]{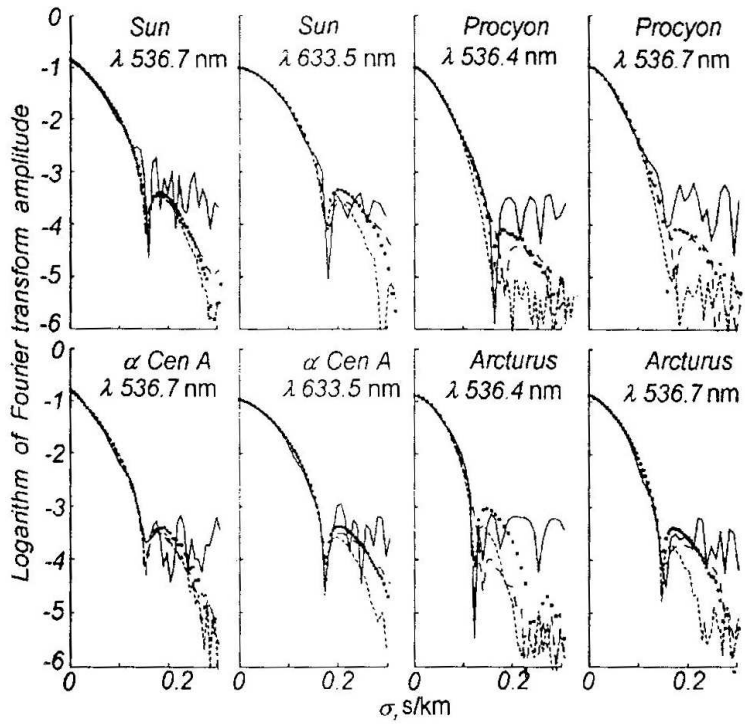}
 \hfill
 \parbox[b]{5.5cm}
{\vspace{1.1cm}
 \caption
{Fourier transforms of the line profiles observed (solid lines)
and calculated with different turbulent velocity distributions:
isotropic micro-macroturbulent velocities (short dashes);
isotropic microturbulence and anisotropic radial-tangential
macroturbulence (small squares); anisotropic radial-tangential
micro-macroturbulent velocities (long dashes). }} \label{F-1}
 \end{figure}

It follows from the comparison of the Fourier transforms of the
observed and calculated line profiles that both microturbulence
models (isotropic and anisotropic) give the same results for the
Sun and $\alpha$ Cen A (Fig. 1). The Fourier transforms for
these stars are more sensitive to macroturbulent velocity
distributions. Their side lobes display some differences for the
isotropic and anisotropic macroturbulence, and we may conclude
that the anisotropic macroturbulence model gives a better
result. For the atmospheres of Procyon and Arcturus (Fig. 1), the
isotropic microturbulence describes line profiles better than
the anisotropic one. The distribution of macroturbulent
velocities remains anisotropic.

  \begin{figure}
 \includegraphics [scale=1.2]{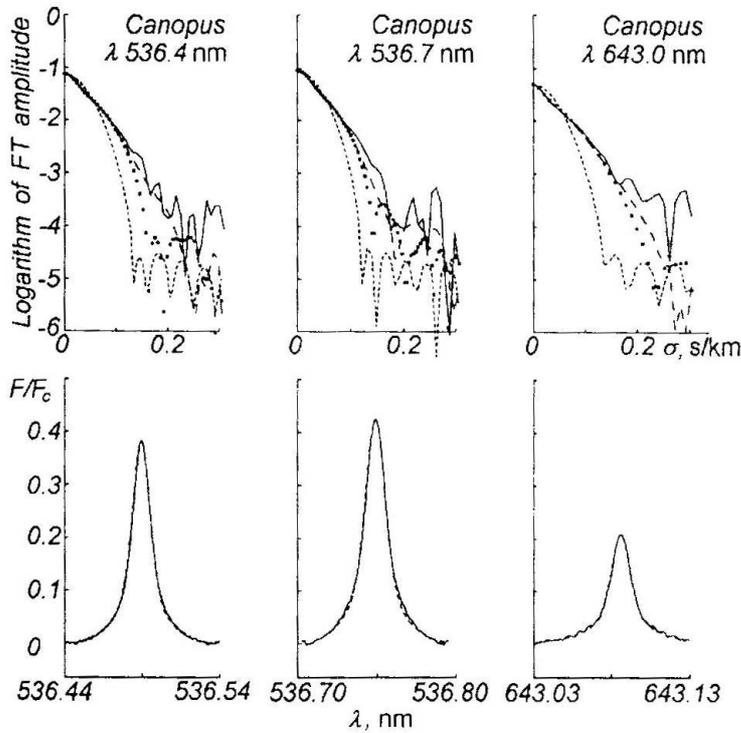}
 \hfill
  \parbox[b]{5.5cm}{\vspace{1.1cm}
 \caption
{The same as in Fig. 1 for Canopus (upper panel). Given on the lower panel
 are line profiles for Canopus: observed (solid lines) and
calculated (dashed lines). }} \label{F-2}
 \end{figure}
Selection of velocity field parameters for Canopus turned out to
be a difficult task. This is so for microturbulent velocities
because the lines in the spectrum of this hot star are weak and
their Fourier transforms have smaller amplitudes. Besides, the
line profiles differ from the profiles in other stars in that
the ratio between wing width and core height is larger: the
lines in Canopus have narrow cores and extended wings, as a
rule. We attempted to describe them with anisotropic velocity
models. When comparing the thermal + microturbulent profile with
observations, we could not determine reliably the microturbulent
velocity because the side lobe was greatly distorted by noises
and its position could not be fixed exactly. The most adequate
value was found to be $V_{mi}^{RT} = 2.1$~km/s. Our calculations
suggested that a rotation velocity estimate of 15 km/s derived
in \cite{47} was inconsistent with the observed profiles and their
Fourier transforms at any microturbulent velocity. For three out
of six selected lines ($\lambda\lambda 536.4$, 536.7, 643.0~nm)
we could fit the synthesized symmetrical line profiles and
Fourier transforms to the observed ones (Fig. 2) only when we
adopted $3.5 \pm 0.2$~km/s for $V_e\sin i$ and the
macroturbulence was assumed to have a large radial component
($17 \pm 2$~km/s) and a relatively small tangential component
($1.3 \pm 1.0$~km/s).

The Fourier transforms of rotation functions plotted in Fig. 3
were calculated for $V_e\sin i$ equal to 3, 3.5, and 4~km/s with
the limb darkening $\varepsilon = 0.6$; the Fourier transform of
the residual observed profile contains the rotation function
only (plus measurement noises). It should be noted that the
other selected lines, for which we could not choose velocity
parameters, are formed higher in the photosphere and they are
likely to be affected by the NLTE effects to a greater extent.
The velocities derived for Canopus do not defy common sense.
Suffice it to look at the profile shape in order to understand
that the rotation velocity cannot be too large, otherwise the
line cores would be bell-shaped like in the line profiles of
Sirius, whose rotation velocity is 15~km/s  \cite{18}. The extended
wings may be produced by radial high-velocity flows that ascend
and then descend without spreading.
 \begin{figure}
 \hspace{0.5cm}
 \includegraphics [scale=1.2]{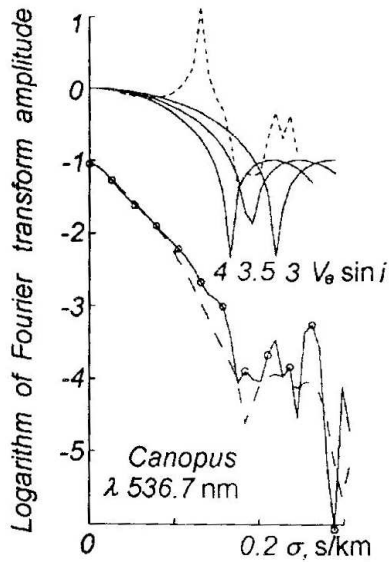}
 \hfill
 \parbox[b]{8.cm}{\vspace{1.1cm}
 \caption
{The Fourier transform of the observed line profile
$d(\sigma)$ (solid line + circles) and the calculated line profile
$f(\sigma)i(\sigma)g(\sigma)$ (long dashes). The Fourier transform
of rotation function together with
observation noises $g(\sigma)n(\sigma)$ (short dashes) derived
from the observed line profile
$d(\sigma)$. Solid lines represent the
Fourier transforms of the rotation functions calculated with
 $V_e\sin i = 3$, 3.5, 4~km/s.  }} \label{F-3}
 \end{figure}

So, the microturbulence in cool solar-type stars has an
isotropic Gaussian distribution over the stellar disk. The
macroturbulence distribution is anisotropic. For hot stars like
Canopus a distinct anisotropy is observed in the distributions
of micro-macroturbulent velocities. When the macroturbulent
velocities determined by different methods and with different
approximations (isotropic and anisotropic) are reduced to a
common system of $rms$ velocities (see $V_{rms}$ in Table 3),
the resulting $V_{rms}$ values are in good agreement. This does
not hold for Canopus, as $V_{ma}^{rad}$ dominates over
$V_{ma}^{tan}$ in it.

\section{Errors of the method}
     \label{Errors}
Two different models of Arcturus  \cite{35} and  \cite{38} give the
differences $\Delta V_{mi} = 0.2$~km/s in microvelocities and
$\Delta V_{ma}= 0.1$~km/s in macrovelocities. For the Sun as a
star we obtained $\Delta V_{mi} = -0.1$~km/s and $\Delta V_{ma}
= 0.1$~km/s from the models HOLMU  \cite{33} and KURUCZ~ \cite{36}. We may
assume, therefore, that the errors due to uncertainties in the
parameters of stellar atmospheres are 0.1--0.2~km/s. When the
rotation velocity of Arcturus was increased by 0.4~km/s, we got
$\Delta V_{mi} = -0.1$~km/s and $\Delta V_{ma} = -0.3$~km/s. The
effect of uncertainty in the damping constant is much weaker for
Procyon, Arcturus, and Canopus than for the Sun and $\alpha$ Cen
A. An increase by a factor of 1.5 in the damping constant gives
$\Delta V_{mi} =-0.1$~km/s and $\Delta V_{ma} =-0.4$~km/s for
the Sun. Errors due to the choice of the local continuum are
also possible. When the continuum for the $\lambda$~536.7~nm
line in the spectrum of Arcturus is drawn 1 percent higher, the
principal lobe in the Fourier transform becomes less steep at
higher frequencies and the micro-macrovelocities diminish by 0.1
and 0.3~km/s. Finally, the velocities found from different lines
(for all stars except Canopus) differ by 0.2 km/s on the average
(Table 3). The difference is due to the fact that the lines form
at different heights as well as to noises in the side lobe in
the Fourier transforms of the observed profiles.

\section{Discussion}
     \label{Discussion}
Table 4 gives final results for the stars studied.

{\bf Microturbulence} varies nonmonotonically with effective
temperature. It is minimum for the Sun and $\alpha$ Cen A and
grows with decreasing and increasing $T_{eff}$. Such a behavior
of microturbulent velocities is well known  \cite{5}.

%
 \begin{table}[!htb] \centering
 \parbox[b]{15cm}{
 \caption
{Velocity field parameters for five stars from the data of
various authors. Type of micro-macro turbulent velocity model is
indicated in the last column \label{T:4} } \vspace{0.3cm}}
 \footnotesize
\begin{tabular}{|cccc|}
  \hline
$V_{mi}$, km/s  & $V_{ma}$, km/s &  Reference &  Note\\
  \hline
 &&           Sun&                                                                                 \\
 0.5$\pm$0.1   &  3.1$\pm$0.1  &    \cite{22} &   $V_{mi}$, $V_{ma}^{RT}$, 5 $< W < 15$~pm, Fourier analysis \\
  --             & 3.8$\pm$0.2   &   \cite{22} & $V_{ma}^{RT}$, $W < 5$~pm, Fourier analysis                      \\
 1.2   &  2.3  &    \cite{9} &$V_{mi}$, $V_{ma}$, 6 $<W<9$~pm                                               \\
 1.08--0.5 &1.92--1.46 &   \cite{3}& $V_{mi}$, $V_{ma}$, $H= 150$--300~km                                 \\
 1.11--1.09& -- & \cite{21}& $V_{mi}$, $\log \tau_5=-1.71\div-3.17$                                       \\
 0.5$\pm$ 0.2& 2.3$\pm$0.4&  \cite{46}& $V_{mi}$, $V_{ma}^{RT}$, 10 $<W<20$~pm                             \\
  -- &4.0&  \cite{46}& $V_{ma}^{RT}$, $W < 10$~pm                                                         \\
  0.8$\pm$0.1 &2.6$\pm$ 0.3&     Present study& $V_{mi}$, $V_{ma}$, $\log \tau_5=- 1.9\pm0.2$       \\
 \hline
&&            a Cen A&
\\
 1.54$\pm$0.08  & -- &   \cite{44}&    $V_{mi}$, Ca I                                                     \\
 1.5--1.06 &   --   & \cite{21}& $V_{mi}$, $\log \tau_5=-1.78\div-2.94$                                   \\
 0.8$\pm$0.2& 2.9$\pm$0.4 &    Present study& $V_{mi}$, $V_{ma}^{RT}$, $\log \tau_5=-2.3\pm 0.4$     \\
 \hline
 &&            Procyon&                                                                             \\
  -- &  7$\pm$0.1 &   \cite{24}  &  $V_{ma}^{RT}$, Fourier analysis                                       \\
 2.1$\pm$0.3 & 4.2$\pm$0.5 & \cite{45}  &  $V_{mi}$, $V_{ma}$                                              \\
 1.6--1.9 &    4.1--3.6 & \cite{3}& $V_{mi}$, $V_{ma}$, $H= 500$--900~km                                  \\
 0.8$\pm$0.3 & 5.9$\pm$0.2& Present study& $V_{mi}$, $V_{ma}^{RT}$, $\log \tau_5= - 1.8\pm0.2$                       \\
 \hline
 &&            Arcturus &                                                                            \\
 1.9 &    2.5--3.5 &     \cite{43} &   $V_{mi}$,  $V_{ma}^{RT}$, Fourier analysis                          \\
 1.7 &4.6&      \cite{28}   & $V_{mi}$, $V_{ma}^{RT}$, Fourier analysis                                     \\
 1.8 & 4.8--5.2&  \cite{26}  &  $V_{mi}$, $V_{ma}^{RT}$, Fourier analysis                                 \\
 1.68--1.87&  -- & \cite{21}& $V_{mi}$, $\log \tau_5= -1.55\div-2.68$                                       \\
 1.0$\pm$0.2& 4.6$\pm$0.3& Present study& $V_{mi}$, $V_{ma}^{RT}$, $\log \tau_5=-1.7\pm0.1$             \\
 \hline
   &&          Canopus&                                                                               \\
   4.5  &   -- &    \cite{2}  &   $V_{mi}$, 1 $ > W>30$ pm, Fe I                                          \\
   6.0  &  -- &  \cite{2}& $V_{mi}$, $1 >W>30$ pm, Fe II, Ti II, Cr II                                      \\
    2.0--4.3 &    -- &  \cite{21} & $V_{mi}$, $\log \tau_5=-1.4\div-2.5$, Fe I                               \\
    2.4--5.4 &    --  &  \cite{21}& $V_{mi}$, $\log \tau_5=-1.3\div-2.7$, Fe II                             \\
   2.1  &   17--1.3 &  Present study &$V_{mi}^{RT}$, $V_{ma}^{rad}$-- $V_{ma}^{tan}$, $\log \tau_5=-1.4\pm0.1$\\
   \hline
\end{tabular}
\end{table}

The height distribution of $V_{mi}$ is evident from studies  \cite{21}
and  \cite{3}: microturbulence decreases with height only in the
atmospheres of the Sun and $\alpha$ Cen A. In Canopus and
Arcturus it increases with height in the atmosphere, which is in
good agreement with the results of other authors  \cite{5}. The
behavior of $V_{mi}$ with depth is peculiar in Canopus  \cite{2,21}:
it drastically increases with height, and the values derived
from the lines of neutral and ionized iron differ considerably.
These peculiarities are typical of F supergiants  \cite{8}. The
simplest isotropic microturbulence model proved to be adequate
for all stars except Canopus; the anisotropic model was used for
that star.

{\bf Macroturbulence} also varies nonmonotonically with
$T_{eff}$. The velocity $V_{ma}$ is minimum for G stars. This is in
conformity with studies  \cite{25,30}, where the temperature
dependence of granulation velocities was determined separately
for stars of luminosity classes III and IV--V. The depth
dependence was determined reliably for the Sun in our studies
 \cite{3,32} --- macroturbulence decreases with height in the
photosphere. Takeda  \cite{46} confirmed this fact from his analysis
of a large number of lines by the new method of multiparametric
fit of line profiles. The same behavior is displayed by $V_{ma}$
in the photosphere of Procyon  \cite{3}.

The isotropic macroturbulence model was found to be quite
adequate for the Sun and $\alpha$~Cen~A. Much better results are
obtained for Arcturus and Procyon with the anisotropic model
with similar radial and tangential macrovelocity components. A
special variant of the anisotropic macroturbulence model with a
sharp asymmetry between the radial and tangential components was
used for the supergiant Canopus. In general, the anisotropic
macroturbulence model may be considered as preferable, since it
is adequate for all stars.

The results obtained for the photospheric velocity field in
Procyon, Sun, $\alpha$ Cen A, and Arcturus are in accord with
the estimates derived from line asymmetries  \cite{25,30}; they are
likely to reflect distinctions in the behavior of the
photospheric overshooting convection in the main sequence stars
and stars on the giant branch  \cite{41}. We discuss below some
special features of these distinctions.

{\bf The Sun as a star.} We emphasize a good agreement between
our results and the data of other authors (Table 4).

One can see from the radial and tangential microturbulence
components determined from the observed center-to-limb
broadening of spectral lines  \cite{3,32,10} that the horizontal
micro-velocity is systematically higher by 1 km/s than the
vertical one and this difference changes slightly with height.
This is confirmed qualitatively by direct stratospheric
observations of solar granulation  \cite{39}. Thus, the center-to-limb
observations of line broadening do not suggest that the
microturbulent velocity is isotropic in the solar photosphere.
We can use, nevertheless, the isotropic model for the Sun as a
star along with the anisotropic model because the relative
contribution from the limb regions decreases and the ratio
between the tangential and radial components is nearly constant.
It is obvious that such an isotropic microvelocity should be
considered as some effective quantity.

It follows from the analysis of line profiles  \cite{3,32,10} and the
direct estimates of $rms$ velocities from Doppler line shifts in
the granulation field  \cite{39} that the anisotropy between the
horizontal and vertical velocities (the predominance of
horizontal velocities) is well defined in large-scale
structures; it is reproduced in 2D and 3D solar granulation
models  \cite{11}. Stratospheric observations  \cite{39} indicate that the
anisotropy is much stronger for the large-scale structures ($L >
3.7^{\prime\prime} \approx 2700$~km) than for the small-scale
ones in the lower and middle photosphere. This may be the cause
why the radial-tangential model of macrovelocities differs
essentially from the isotropic model and is more adequate for
the observed profiles.

The behavior of the parameters found for the photospheric
velocity field reflects a decrease of overshooting convection
velocities as the predominant process, with oscillatory motions
being the governing factor. Multidimensional hydrodynamic
simulations and observations of granulation with high spatial
resolution  \cite{11,34} suggest that the lower and middle photosphere
regions, where most photospheric lines are formed, are
controlled by overshooting convection, while oscillatory motions
become predominant near the temperature minimum and above it.

{\bf $\alpha$ Centauri A.} The star is very similar to the Sun
in most parameters, but its radius is larger by 23 percent and
the mass by 8.5 percent. The gravitational acceleration is
smaller by 29 percent than on the Sun (Table 1). The abundances
of heavy elements are higher by 0.1--0.3~dex in $\alpha$ Cen A
 \cite{20,21,44}, The microturbulent velocity diminishes with height
 \cite{21}, similar to the Sun, but the amplitudes of
micro-macroturbulent velocities are larger (Tables~3 and 4). We
may assume that higher amplitudes are related to the
above-mentioned peculiarities. Indeed, it follows from 3D HD
models  \cite{41} that the convection velocities, and the atmospheric
horizontal flow velocities in the first place, increase with
diminishing gravitational accelerations in stellar convective
envelopes and atmospheres. The reason is that the pressure and
density height scales grow with decreasing $g$. This results in
a growth of the horizontal size of inhomogeneities, since this
size is proportional to the density height scale; the horizontal
flow velocities increase as well, as it follows from the
condition of conservation of mass. The second effect which
enhances the vertical and horizontal velocities is a decrease in
density and with it in the viscosity of the medium. The third
factor is an elevated metallicity, which results in an elevated
electron density in the lower photosphere and in an enhanced
opacity, as the main opacity source in the continuum is the
negative H$^-$ ions, their concentration being proportional to
electron density.

{\bf Procyon} has an effective temperature 700-750 K higher than
that of the Sun and a gravitational acceleration 60 percent
lower. The radiation dynamics conditions in the upper layers of
the star were studied in detail in  \cite{11} and  \cite{41} based on 3D
models. Their major peculiarity is that the thermal convection
on Procyon, as distinct from the Sun, makes its way into
photospheric layers, since the major supplier of free electrons
in the lower photospheric layers on Procyon, due to their higher
temperature, is the hydrogen abundant there. As a result, the
H$^-$ opacity is much greater, it is extremely sensitive to
temperature variations, thus maintaining adiabatic conditions
for flows in photospheric layers. The artificial granulation
displays, therefore, brightness fluctuations stronger than on
the Sun, their growth toward the limb, and a limb effect inverse
to the solar one  \cite{11}. Macrovelocities on Procyon are by a
factor of 2--3 higher than on the Sun, the cause being the same
as for $\alpha$ Cen A; they are higher also due to the fact that
convective flows have to transport larger energy fluxes.

{\bf Arcturus} is a red metal-poor giant with low-temperature,
low-density atmospheric layers. Considering the results of 3D
simulations of stellar granulation on $\alpha$  Cen B  \cite{41},
which has $T_{eff}$ lower by 600 K than on the Sun, we may
assume that the overshooting convection on Arcturus is shifted
to deeper layers and manifests itself primarily via velocity
field. One more peculiarity is that the size of inhomogeneities
is comparable to the star radius (up to 10 percent of $R_*$)
owing to large height scales of density and pressure. Therefore,
no more than several ``granules'' can be found simultaneously on
the entire apparent disk of the star. The velocities found by us
for Arcturus are higher than on the Sun (Table 3). At the same
time, the semiempirical four-component model by Dravins  \cite{17},
which was specially built for the interpretation of observed
bisectors, gives velocities slightly lower than on the Sun.
These discrepancies seem to be due to the contribution from
oscillations, which we partially detect when analyzing spectral
lines.

{\bf Canopus} is a hot F supergiant distinctive in that its
outer layers are in radiative equilibrium. This is the cause for
the nonstandard results obtained for $ V_{ma}^{RT}$ -- a high
amplitude of the vertical component with no appreciable
horizontal component. Similar solutions were derived from the
semiempirical four-component model of Canopus  \cite{17} and the
analysis of bisectors for the supergiant $\gamma$ Cygni  \cite{31}. A
common peculiarity in the line asymmetry in the stars of this
type is that it is inverse with respect to the classical line
profiles observed in stars with convection-induced photospheric
flows. De Jager  \cite{14} assumed that such a pattern of the
large-scale velocity field in supergiant atmospheres is
conditioned by gravity waves.

\section{Conclusion}
     \label{Conclusion}
The isotropic distribution of microturbulent velocities and the
radial-tangential distribution of macroturbulent velocities seem
to represent most adequately the velocity field in the
atmospheres of the Sun, $\alpha$ Cen A, Procyon, and Arcturus
within the scope of their homogeneous models. For Canopus, the
microturbulence is anisotropic and the macroturbulence displays
a pronounced asymmetry between its radial and tangential
components. The rotation velocity found by us for Canopus turned
out to be low.

\vspace{1.cm}
 {\bf Acknowledgements.} We thank Prof. D. Dravins
for furnishing the observational data. The study was partially
financed by the State Fundamental Research Foundation of Ukraine
(Grant No. 6.4/192).



\end{document}